\documentclass[notoc]{JHEP3}
\usepackage{graphicx}
\usepackage{amsmath}
\usepackage{amsfonts}
\usepackage{amssymb}
\usepackage{epsfig}
\def \als {\alpha_{\mathrm{s}}} 
\def \lQ {\Lambda_{\mathrm{QCD}}}

\def \be {\mathbf{E}}

\def \br {\mathbf{r}}
\def \brg {\mathbf{R}}

\def \cf {C_F}

\def \nc {N_c}

\def \bp {\mathbf{p}}
\def \bpg {\mathbf{P}}

\newcommand{\Tint}[1]{{\hbox{$\sum$}\!\!\!\!\!\!\!\int\,}_{\!\!\!\!\raise-0.9ex\hbox{$\scriptstyle{#1}$}}}

\def \m2   {\mu^{2 \epsilon}}
\def\alVs{\alpha_{V_s}}
\def\alVo{\alpha_{V_o}}
\def\siml{{\ \lower-1.2pt\vbox{\hbox{\rlap{$<$}\lower6pt\vbox{\hbox{$\sim$}}}}\ }}
\def\simg{{\ \lower-1.2pt\vbox{\hbox{\rlap{$>$}\lower6pt\vbox{\hbox{$\sim$}}}}\ }}
\def\bfnabla{\mbox{\boldmath $\nabla$}}

\def\bfsigma{\mbox{\boldmath $\sigma$}}

\def \nn{\nonumber}

\title{The spin-orbit potential and Poincar\'e invariance in finite temperature pNRQCD}

\author{Nora Brambilla\\
Physik-Department, Technische Universit\"at M\"unchen,
James-Franck-Str. 1, 85748 Garching, Germany}
\author{Miguel \'Angel Escobedo\\
Physik-Department, Technische Universit\"at M\"unchen,
James-Franck-Str. 1, 85748 Garching, Germany
}
\author{Jacopo Ghiglieri\\
Physik-Department, Technische Universit\"at M\"unchen,
James-Franck-Str. 1, 85748 Garching, Germany and Excellence Cluster Universe, Technische Universit\"at M\"unchen, 
Boltzmannstr. 2, 85748, Garching, Germany }
\author{Antonio Vairo\\
Physik-Department, Technische Universit\"at M\"unchen,
James-Franck-Str. 1, 85748 Garching, Germany}

\preprint{TUM-EFT 15/10}

\abstract{
Heavy quarkonium at finite temperature has been the subject of intense theoretical studies, 
for it provides a potentially clean probe of the quark-gluon plasma.
Recent studies have made use of effective field theories to exploit in a systematic manner 
the hierarchy of energy scales that characterize the system. 
In the case of a quarkonium in a medium whose temperature 
is smaller than the typical momentum transfer in the bound state but 
larger than its energy, the suitable effective field theory is 
$\mathrm{pNRQCD}_{\rm HTL}$, where degrees of freedom with 
energy or momentum larger than the binding energy have been integrated out.
Thermal effects are expected to break Poincar\'e invariance, which, at zero temperature, 
manifests itself in a set of exact relations between the matching coefficients 
of the effective field theory. In the paper, we evaluate the leading-order thermal 
corrections to the spin-orbit potentials of $\mathrm{pNRQCD}_{\rm HTL}$ and show that 
Poincar\'e invariance is indeed violated.
}


\begin{document}

\section{Introduction}
Heavy quarkonium can be systematically studied by means of non-relativistic 
effective field theories (EFTs) \cite{Brambilla:2004jw}.  
Recently, the non-relativistic EFT framework has been extended to allow 
the study of heavy quarkonium in a thermal bath 
\cite{Laine:2006ns,Laine:2007gj,Burnier:2007qm,Brambilla:2008cx,Escobedo:2008sy,Brambilla:2010vq,Escobedo:2010tu,Escobedo:2011ie}.
The relevance of heavy quarkonium, as a probe of the hot medium created by heavy ion collisions 
at modern accelerator machines, has been remarked since long time \cite{Matsui:1986dk}. 

Heavy quarkonium is characterized by the scales typical of a non-relativistic bound state 
and the thermal bath is characterized by the thermodynamical scales.
The former are the heavy-quark mass, $m$, the inverse of the typical radius of the system, 
$1/r$, and the binding energy, whereas the latter are the
temperature $T$ (or rather multiples of $\pi T$) and possibly smaller scales. 
The non-relativistic scales are hierarchically ordered: the mass is much larger 
than the inverse of the typical radius of the system, which in turn is much larger 
than the binding energy. In the weak-coupling regime, which may be relevant for the lowest 
quarkonium resonances \cite{Brambilla:2004wf,Brambilla:2010cs}, 
the inverse of the typical radius of the system scales like  
$m\als$, while the binding energy scales like $m\als^2$.
In the following, we will assume that the quarkonium is a weakly coupled 
bound state and that the temperature is such that 
\begin{equation}
m\gg m\als\gg T \gg m\als^2\,.
\label{hierarchy}
\end{equation}
Moreover, we will assume that all other thermodynamical scales as well as the typical hadronic scale, 
$\lQ$, are either of the same order as or smaller than the binding energy. As it has been argued 
in \cite{Brambilla:2010vq,Vairo:2010bm}, this situation may be relevant for 
the phenomenology of the ground states of bottomonium in heavy-ion collisions at the LHC.
A detailed study of the spectrum and the width of quarkonium under a more restrictive 
hierarchy than \eqref{hierarchy} can be found in \cite{Brambilla:2010vq}.

One may exploit the hierarchy \eqref{hierarchy} by systematically integrating out degrees 
of freedom associated with the scales $m$, $m\als$ and $T$ and by substituting QCD 
with low-energy EFTs better suited to describe the quarkonium in the assumed regime.
It turns out that, under the scale hierarchy \eqref{hierarchy}, 
quarkonium behaves like a Coulombic bound state with small thermal corrections  
induced by the thermal bath.

After integrating out the scales $m$ and $m\als$ from QCD, one obtains
potential Non-Relativistic QCD (pNRQCD) \cite{Pineda:1997bj,Brambilla:1999xf}. 
Although in a non-relativistic EFT Poincar\'e invariance is not apparent, 
still such invariance is realized non-linearly through a set of relations among its matching 
coefficients \cite{Luke:1992cs,Manohar:1997qy,Brambilla:2001xk,Brambilla:2003nt,Brambilla:2008zg}.
One of these relations is the so-called Gromes relation \cite{Gromes:1984ma} 
that relates the spin-orbit potential with the static potential.

Integrating out the temperature $T$ from pNRQCD, one obtains 
$\mathrm{pNRQCD}_{\rm HTL}$ \cite{Brambilla:2008cx,Vairo:2009ih}. 
This EFT has the same degrees of freedom as pNRQCD, but the 
matching coefficients and, in particular, the potentials get 
thermal corrections, while the Yang--Mills Lagrangian modifies 
into the Hard Thermal Loop (HTL) Lagrangian \cite{Braaten:1991gm}. 
Because the thermal bath introduces a preferred reference 
frame (the one in which the thermal bath is at rest), one expects that 
$\mathrm{pNRQCD}_{\rm HTL}$ breaks Poincar\'e invariance;\footnote{
More precisely, the thermal bath breaks Lorentz invariance. 
Since, however, in the context of non-relativistic EFTs exact relations among 
the matching coefficients have been derived from the Poincar\'e algebra 
\cite{Brambilla:2001xk,Brambilla:2003nt,Brambilla:2008zg}, an approach 
that may be traced back to \cite{Dirac:1949cp}, in the paper we will keep 
referring to the broader Poincar\'e invariance.
} 
this would have consequences for the properties of a heavy quarkonium 
moving with a certain velocity relative to the thermal bath. 
In order to make the statement more quantitative, we calculate in the 
paper the leading-order thermal contributions to the spin-orbit potential 
of $\mathrm{pNRQCD}_{\rm HTL}$ and check them against the Gromes relation. 

The paper is organized as follows. In Sec.~\ref{secpnrqcd}, we write 
the pNRQCD Lagrangian and review how Poincar\'e invariance is realized in
the effective field theory at $T=0$. Section~\ref{secpnrqcdhtl} is devoted to the
computation of the leading thermal corrections to the singlet static potential 
and to the part of the spin-orbit potential of $\mathrm{pNRQCD}_{\rm HTL}$
that depends on the centre-of-mass momentum.
In Sec.~\ref{secpoi}, we show that the thermal corrections violate the Gromes relation.
In Sec.~\ref{secspinorb}, we calculate the leading thermal corrections to the complete 
spin-orbit potential and, in Sec.~\ref{secconclusions}, we present our conclusions.

\section{pNRQCD}
\label{secpnrqcd}
The largest scale in the hierarchy \eqref{hierarchy} is the heavy-quark mass. 
Integrating it out leads to NRQCD \cite{Caswell:1985ui,Bodwin:1994jh}. 
The matching is insensitive to the lower-energy scales, such as the thermal ones,
therefore the EFT Lagrangian is the same as the one at zero temperature.

The next relevant scale is the inverse of the typical radius of the system, which, in the weak-coupling regime, 
scales like $m\als$.  By integrating out this scale, we obtain weakly-coupled pNRQCD  
\cite{Pineda:1997bj,Brambilla:1999xf}, whose Lagrangian reads\footnote{
We adopt here and in the following the notation of \cite{Brambilla:2003nt}.}
\begin{eqnarray}
&& \hspace{-8mm} {\mathcal L}_{\rm pNRQCD} = \int d^3r \, {\rm Tr} \,\left\{
  {\rm S}^\dagger \left( i\partial_0 - h_s \right) {\rm S}
+ {\rm O}^\dagger \left( iD_0 - h_o \right) {\rm O}
\right.
\nn\\
&&\qquad
- \left[ ({\rm S}^\dagger h_{so} {\rm O} + {\rm H.C.}) + {\rm C.C.} \right] \;
- \left[ {\rm O}^\dagger h_{oo} {\rm O} + {\rm C.C.}  \right]  \;
\nn \\
&&\qquad
\left.
 - \left[ {\rm O}^\dagger h_{oo}^A {\rm O}  h_{oo}^B + {\rm C.C.}  \right]
\right\}
- \frac{1}{4} F^a_{\mu \nu} F^{a\,\mu \nu}+ \sum_{i=1}^{n_f}\bar{q}_i\,iD\!\!\!\!/\,q_i\,.
\label{pnrqcd}
\end{eqnarray}
The fields $\mathrm{S}=S\,\mathbf{1}_c/\sqrt{N_c}$ and
$\mathrm{O}=O^a\,T^a/\sqrt{T_F}$, are the quark-antiquark colour-singlet and
colour-octet fields respectively, $n_f$ is the number of light quark fields, $q_i$,
$N_c=3$ is the number of colours, $T_F=1/2$ and 
$iD_0 \mathrm{O}=i\partial_0 \mathrm{O}-gA_0 \mathrm{O} + \mathrm{O} gA_0$. 
The trace is intended over colour and spin indices; C.C. stands
for charge conjugation and H.C. stands for Hermitian conjugation. 
Quark-antiquark fields depend on the centre-of-mass coordinate,  
$\brg$, on the relative distance between the quark and the antiquark, $\br$, and on time.
Gluon fields depend only on $\brg$ and on time; this is achieved by multipole expanding them 
in $\br$.  The operators $h_s$ and $h_o$ do not contain gluon fields, except for the case 
of $h_o$ in covariant derivatives, and may be interpreted 
as the colour-singlet and the colour-octet Hamiltonians respectively; 
terms contributing to $h_s$ and $h_o$ are ordered in powers of $1/m$. 
The operators $h_{so}$, $h_{oo}$, $h_{oo}^A$ and $h_{oo}^B$ contain gluon fields; 
terms contributing to them are ordered in powers of $1/m$ and $\br$.
We will detail these terms in the following. 
Since, again, the matching is insensitive to the lower-energy scales, 
the Lagrangian \eqref{pnrqcd} is the same as the one at zero temperature.
We assume to be in the laboratory reference frame, which we define as the frame 
where an infinitely heavy quarkonium would be at rest.

The singlet and octet Hamiltonians read
\begin{equation}
h_{s,o} = \frac{{\bf p}^2}{m} + \frac{{\bf P}^2}{4m}
+ V_{s,o}^{(0)} + \frac{V^{(1)}_{s,o}}{m} + \frac{V^{(2)}_{s,o}}{m^2}+\ldots\,,
\label{hamil}
\end{equation}
where $m$ is the heavy-quark mass, ${\bf P} = -i{\bf D}_\brg$ and
${\bf p} = -i\bfnabla_\br$. Terms in the Hamiltonian have been ordered 
in powers of $1/m$; the dots stand for higher-order terms.
We recall that the non-relativistic power counting has  
$1/r \sim p\sim m\als$ and $V_{s,o}^{(0)} \sim m\als^2$, while the centre-of-mass 
momentum, $P$, may be as large as $T$. 

The static potentials read
\begin{equation}
V^{(0)}_s(r)=-C_F\frac{\alVs}{r}\,,\qquad V^{(0)}_o(r)=\frac{1}{2N_c}\frac{\alVo}{r}\,,
\label{staticpot}
\end{equation}
where $C_F=(N_c^2-1)/(2N_c)$ and $\alVs$ and $\alVo$ are series in
$\als$; $\alVs$ is known up to three loops \cite{Smirnov:2009fh,Anzai:2009tm}, 
whereas $\alVo$ up to two loops \cite{Kniehl:2004rk}. 
At leading order, it holds that $\alVs=\alVo=\als$.

The singlet and octet propagators of pNRQCD can be expanded as 
\begin{eqnarray}
\hspace{-8mm} 
S^{\rm singlet}(E) =\frac{i}{E -h_s - \Sigma_s(E) +i\eta} &=& \frac{i}{E -h_s+i\eta} 
\nonumber\\
& & \hspace{-10mm}
+ \frac{i}{E -h_s+i\eta} \Sigma_s(E) \frac{1}{E -h_s+i\eta} +\ldots,
\label{hqprops}
\\
\hspace{-8mm} 
S^{\rm octet}_{ab}(E) =\left( \frac{i}{E -h_o-\Sigma_o(E)+i\eta} \right)_{ab} &=& \frac{i\delta_{ab}}{E -h_o+i\eta} 
\nonumber\\
& & \hspace{-10mm}
+ \frac{i}{E -h_o+i\eta} \Sigma_o(E)_{ab} \frac{1}{E -h_o+i\eta} +\ldots,
\label{hqpropo}
\end{eqnarray}
where $\Sigma_{s,o}$ are the colour-singlet and colour-octet self-energies.
According to the power counting of pNRQCD and the fact that $V^{(1)}_{s,o} \sim V^{(2)}_{s,o} \ll m \als^2$, 
the propagator $1/(E -h_{s,o}+i\eta)$ may, in turn, be expanded as 
\begin{eqnarray}
\hspace{-8mm} 
\frac{1}{E -h_{s,o} +i\eta} &=& \frac{1}{E -h_{s,o}^{(0)}+i\eta} 
\nonumber\\
&& \hspace{-10mm}
+ \frac{1}{E -h_{s,o}^{(0)}+i\eta} 
\left[ \frac{{\bf P}^2}{4m} + \frac{V^{(1)}_{s,o}}{m} + \frac{V^{(2)}_{s,o}}{m^2} +\ldots \right]
\frac{1}{E -h_{s,o}^{(0)}+i\eta} +\ldots,
\label{hqpropsozero}
\end{eqnarray}
where $\displaystyle h_s^{(0)}=\frac{\bp^2}{m}-\cf\frac{\als}{r}$ and 
$\displaystyle h_o^{(0)}=\frac{\bp^2}{m}+\frac{1}{2\nc}\frac{\als}{r}$ 
are the leading-order singlet and octet Hamiltonians respectively. 

The explicit form of the non-static potentials $V^{(1)}_s$ and $V^{(2)}_s$ can be read from
\cite{Brambilla:1999xj,Brambilla:2004jw}. In particular, we will
concern ourselves with the colour-singlet, $V_{LS\,s}$, and 
the colour octet, $V_{LS\,o}$, spin-orbit potentials, which
are part of $V^{(2)}_s$ and $V^{(2)}_o$ respectively. 
They can be conveniently split into a part that
depends on the centre-of-mass momentum $\bpg$ and a part that depends
on the relative momentum $\bp$:
\begin{eqnarray}
\hspace{-8mm} 
V_{LS\,s}= \frac{({\bf r}\times {\bf P}) \cdot (\bfsigma^{(1)} - \bfsigma^{(2)})}{4 m^2} V_{LS\,s a}(r)
+ \frac{({\bf r}\times {\bf p}) \cdot (\bfsigma^{(1)} + \bfsigma^{(2)})}{2m^2} V_{LS\,s b}(r),
\label{defvls}
\\
\hspace{-8mm} 
V_{LS\,o}= \frac{({\bf r}\times {\bf P}) \cdot (\bfsigma^{(1)} - \bfsigma^{(2)})}{4 m^2} V_{LS\,o a}(r)
+ \frac{({\bf r}\times {\bf p}) \cdot (\bfsigma^{(1)} + \bfsigma^{(2)})}{2m^2} V_{LS\,o b}(r),
\label{defvlso}
\end{eqnarray}
where the Pauli matrices $\bfsigma^{(1)}$ and $\bfsigma^{(2)}$ act on the heavy quark 
and antiquark respectively. At leading order, it holds that
\begin{eqnarray}
V_{LS\,s a}(r) \! = \! -\frac{\cf}{2}\frac{\als}{r^3},\quad
V_{LS\,s b}(r) \! = \! \frac{3\cf}{2}\frac{\als}{r^3},\quad 
V_{LS\,o a}(r) \! = \! \frac{1}{4\nc}\frac{\als}{r^3},\quad
V_{LS\,o b}(r) \! = \! -\frac{3}{4\nc}\frac{\als}{r^3},
\nn
\\
\label{spinorbitlo}
\end{eqnarray}
which implies that $V_{LS\,s a}\sim V_{LS\,s b}\sim V_{LS\,o a}\sim V_{LS\,o b} \sim m^3\als^4$.

The matching coefficients, and among them the potentials, that 
appear in the pNRQCD Lagrangian obey a set of relations due to Poincar\'e
invariance \cite{Brambilla:2003nt}. For the kinetic terms in Eq. \eqref{hamil}, 
such relations impose that the coefficient of the operator ${{\bf P}^2}/{(4m)}$ is one, 
which also implies that the coefficient of the operator ${{\bf p}^2}/{m}$ is one.
Among the relations fulfilled by the potentials, there are some exact relations linking the 
spin-orbit potentials $V_{LS\,s a}$ and $V_{LS\,o a}$ to the static potentials:
\begin{equation}
V_{LS\,sa}(r)=-\frac{V_s^{(0)}(r)^\prime}{2r}\,,\qquad  V_{LS\,oa}(r)=-\frac{V_o^{(0)}(r)^\prime}{2r}\,,
\label{gromes}
\end{equation}
where $f(r)'\equiv df(r)/dr$. The first relation is known as the Gromes relation, because 
it was first derived in \cite{Gromes:1984ma} from the transformation properties 
of some Wilson loops under Lorentz boosts (see also \cite{Brambilla:2001xk}).

In the next section, we will compute the leading thermal correction to the spin-orbit 
potential $V_{LS\,sa}$ by matching to ${\rm pNRQCD_{\rm HTL}}$. To this end we will need some 
of the terms appearing in the operator $h_{so}$ of the pNRQCD Lagrangian \eqref{pnrqcd}.
These may be ordered in powers of $1/m$ and $\br$ as 
\begin{equation} 
h_{so} = h_{so}^{(0,1)} + h_{so}^{(0,2)} + h_{so}^{(1,0)} + h_{so}^{(1,1)} + h_{so}^{(2,0)}+\ldots\,,
\end{equation}
where the indices $(i,j)$ refer to the order in powers of $1/m$ and $\br$ respectively. 
The dots stand for higher-orders terms.
The explicit expressions of $ h_{so}^{(i,j)}$ may be taken from \cite{Brambilla:2003nt} and read
\begin{eqnarray}
&&h_{so}^{(0,1)} = - \frac{V_{so}^{(0,1)}(r)}{2} {\bf r}\cdot g {\bf E}\,,
\\
&&h_{so}^{(1,0)} = - \frac{c_F}{2m} \, V_{so\,b}^{(1,0)}(r) \, \bfsigma^{(1)}\cdot g {\bf B}
\nn\\
\label{1/m}
&&
\qquad-\frac{1}{2m} \, \frac{V_{so\,c}^{(1,0)}(r)}{r^2}
\, ({\bf r}\cdot\bfsigma^{(1)})\, ({\bf r}\cdot g {\bf B})
- \frac{1}{m}\frac{V_{so\,d}^{(1,0)}(r)}{2 r} {\bf r}\cdot g {\bf E}\,,
\\
&&h_{so}^{(1,1)}  = \frac{1}{8m} \, V_{so}^{(1,1)}(r) \,
\{ {\bf P} \cdot ,{\bf r}\times g {\bf B} \} +\ldots\,,
\label{1/m1m2}
\end{eqnarray}
\begin{eqnarray}
&&h_{so}^{(2,0)} = \frac{c_s}{16m^2} V_{so\,a}^{(2,0)}(r) \, \bfsigma^{(1)}\cdot [{\bf P} \times, g {\bf E}]
\nn\\
&&\qquad
+ \frac{1}{16m^2} \frac{V_{so\,b'}^{(2,0)}(r)}{r^2} \,
({\bf r}\cdot\bfsigma^{(1)}) \,
\{ {\bf P} \cdot, (g {\bf E} \times {\bf r}) \}
\nn\\
&&\qquad
+ \frac{1}{16m^2} \frac{V_{so\,b''}^{(2,0)}(r)}{r^2} \,
\{ ({\bf r}\cdot g {\bf E}),
{\bf P} \cdot ({\bf r} \times \bfsigma^{(1)})\}
\nn\\
&&\qquad
+ \frac{1}{16m^2} \frac{V_{so\,b'''}^{(2,0)}(r)}{r^2} \,
\{({\bf r}\cdot{\bf P}),
\bfsigma^{(1)} \cdot ({\bf r} \times g {\bf E}) \}
\nn\\
\label{1/m2}
&&\qquad
+ \frac{1}{8m^2} \, \frac{V_{so\, e}^{(2,0)}(r)}{r}\,
\{ {\bf P} \cdot ,{\bf r}\times g {\bf B} \} +\ldots\,.
\end{eqnarray}
Charge conjugation invariance requires that $h_{so}^{(0,2)} = 0$.
The field $\be$ is the chromoelectric field, $E^i=F^{i0}$,  
the field ${\bf B}$ is the chromomagnetic field, $B^i=-\varepsilon_{ijk}F^{jk}/2$, 
$[\bpg\times,g\be]=\bpg\times g\be-g\be\times\bpg$ and similarly for the anticommutators. 
For $h^{(1,1)}_{so}$ and $h^{(2,0)}_{so}$ only the ${\bf P}$-dependent terms have been displayed.
The coefficients $c_F$ and $c_s$ are inherited from NRQCD and encode non-analytical 
contributions in $1/m$, whereas the various $V_{so}^{(i,j)}(r)$ come from the matching to pNRQCD and 
encode non-analytical contributions in $r$. 
At leading order in the coupling, the matching gives $c_F=c_s=1$ and
$V_{so}^{(0,1)}(r) = V_{so\,b}^{(1,0)}(r) = V_{so\,a}^{(2,0)}(r) = V_{so}^{(1,1)}(r) = 1$, 
while all other matching coefficients are of order $\als$ or smaller.

Poincar\'e invariance imposes further constraints on the matching coefficients.
For instance, at the NRQCD level, we have that $2c_F-c_s-1=0$ to all orders \cite{Luke:1992cs,Manohar:1997qy}.
At the level of pNRQCD, the following exact relations hold \cite{Brambilla:2003nt} 
\begin{eqnarray}
&&  V_{so}^{(1,1)}(r) = V_{so}^{(0,1)}(r)\,,
\label{poincare11}
\\
&&2\, c_F V_{so\,b}^{(1,0)}(r) - c_s V_{so\,a}^{(2,0)}(r) = V_{so}^{(0,1)}(r) \,, 
\label{gromesmatching0}
\\
&&2\, c_F V_{so\,b}^{(1,0)}(r) - c_s V_{so\,a}^{(2,0)}(r) - V_{so\,b''}^{(2,0)}(r) = \left(r \, V_{so}^{(0,1)}(r)\right)^\prime\,.
\label{gromesmatching}
\end{eqnarray}
Combining the last two it follows that 
\begin{equation}
V_{so\,b''}^{(2,0)}(r) = -r V_{so}^{(0,1)}(r)^\prime\,.
\label{gromesmatching1}
\end{equation}
An interesting consequence of this relation is that, since $V_{so}^{(0,1)}(r)$ is at least 
of order $\als^2$ \cite{Brambilla:2006wp} but has not infrared divergences at that order 
\cite{Brambilla:2009bi}, $V_{so\,b''}^{(2,0)}(r)$ is at least of order $\als^3$.

\section{pNRQCD$_{\rm HTL}$}
\label{secpnrqcdhtl}
In this section, we compute the leading temperature-dependent correction to the 
colour-singlet static potential, $V^{(0)}_s$, and spin-orbit potential, $V_{LS\,s a}$, which 
amounts to matching the  colour-singlet static and spin-orbit potentials in the EFT that follows 
from pNRQCD after integrating out the temperature.
This EFT is called  $\textrm{pNRQCD}_\mathrm{HTL}$  \cite{Brambilla:2008cx,Escobedo:2008sy,Vairo:2009ih}. 
Its Lagrangian reads
\begin{eqnarray}
{\mathcal L}_{\textrm{pNRQCD}_\mathrm{HTL}} &=& 
 \int d^3r \; {\rm Tr} \,  
\left\{ {\rm S}^\dagger \left[ i\partial_0 - h_s - \delta V_s \right] {\rm S} 
+ {\rm O}^\dagger \left[ iD_0 -h_o - \delta V_o \right] {\rm O} \right\}
\nonumber\\
&&
+{\mathcal L}_{\rm HTL}+\ldots\,,
\label{pNRQCDHTL}	
\end{eqnarray}
where $\delta V_s$ and $\delta V_o$ are the thermal corrections 
to the colour-singlet and colour-octet potentials respectively, 
${\mathcal L}_{\rm HTL}$ is the HTL Lagrangian for the gauge and 
light-quark degrees of freedom \cite{Braaten:1991gm} and the dots stand 
for singlet-octet and octet-octet operators suppressed by powers 
of $1/m$ and $\br$, whose computation is beyond the scope of the paper.

The terms $\delta V_s$ and $\delta V_o$ encode the corrections 
to the potentials induced by the thermal bath. They are evaluated by matching 
real-time propagators in pNRQCD with the corresponding expressions in $\textrm{pNRQCD}_\mathrm{HTL}$.
It has been pointed out in \cite{Brambilla:2008cx,Brambilla:2010vq} that the doubling 
of degrees of freedom, typical of the real-time formalism, does not affect heavy quarks
for which the unphysical degrees of freedom decouple. Hence, in the following, 
we will only deal with the physical singlet and octet propagators, which have been 
written in Eqs. \eqref{hqprops} and  \eqref{hqpropo} respectively.
The matching may be done perturbatively, because we have assumed $T \gg \lQ$;
contributions coming from other thermal scales do not affect the potentials, 
because we have assumed that the other thermal scales are at most as large as 
the binding energy. More specifically, the pNRQCD singlet propagator \eqref{hqprops} is matched to 
the $\textrm{pNRQCD}_\mathrm{HTL}$ expression
\begin{eqnarray}
&& \hspace{-5mm}
Z_s^{1/2} \frac{i}{E -h_s -\delta V_s  + i\eta} Z_s^{1/2\,\dagger} 
= \frac{i}{E -h_s + i\eta} 
\nonumber\\
&&
\hspace{12mm}
+ \frac{i}{E -h_s + i\eta} \delta V_s \frac{1}{E -h_s + i\eta} 
+ \left\{ \delta Z_s , \frac{i}{E -h_s + i\eta} \right\} + \dots\,.
\label{matchingHTL}
\end{eqnarray}
There is no self-energy contribution in \eqref{matchingHTL}, 
because this would correspond to a scaleless integral eventually irrelevant 
(e.g. in dimensional regularization it would vanish).
$Z_s^{1/2} = 1 + \delta Z_s$ is the normalization of the singlet field in $\textrm{pNRQCD}_\mathrm{HTL}$;
$\delta Z_s$ is of order $\als$ and a function of $\br$, which implies that it does not 
commute with $h_s$. At our accuracy, $\delta Z_s$ is real.

We will assume the thermal bath to be a quark-gluon plasma at rest with respect to the 
laboratory reference frame. This implies, in particular, that the Bose--Einstein distribution, 
\begin{equation}
n_\mathrm{B}(x)=\frac{1}{e^{x/T}-1}\,,
\end{equation}
which describes the distribution of the gluons in the bath, depends only on their energy.

\FIGURE[ht]{
\parbox{15cm}{
\centering
\includegraphics[width=7cm]{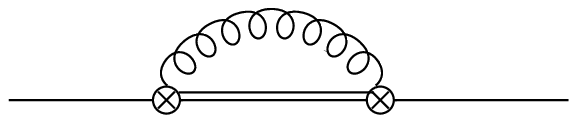}
\caption{The leading heavy quarkonium self-energy diagram. 
Single lines stand for leading-order quark-antiquark colour-singlet propagators, $i/(E -h_s^{(0)} + i\eta)$,
double lines for leading-order quark-antiquark colour-octet propagators, $i\delta_{ab}/(E -h_o^{(0)}+i\eta)$,  
curly lines for gluons and the vertices are the chromoelectric dipole vertices induced by $h_{so}^{(0,1)}$.}
\label{fig:static}}
}

\subsection{Singlet static potential in $\mathrm{pNRQCD}_{\rm HTL}$}
We start by evaluating the leading thermal correction of the static potential, 
$\delta V_s^{(0)}$, which was first done in  \cite{Brambilla:2008cx}, and the 
leading thermal correction of $\delta Z_s$, which is new.
These corrections originate from the pNRQCD diagram shown in Fig.~\ref{fig:static}, 
whose contribution to $\Sigma_s(E)$ reads 
\begin{eqnarray}
\nonumber
&& \hspace{-8mm}
\Sigma_s(E)^{\hbox{\tiny Fig.\,\ref{fig:static}}}=
-iC_F g^2 \left(V_{so}^{(0,1)}(r)\right)^2\frac{2}{3} r^i
\int\frac{\,d^4k}{(2\pi)^4}\frac{i}{E-h_o^{(0)}-k_0+i\eta}k_0^2
\\
&&\hspace{3.5cm}\times\left[\frac{i}{k_0^2-k^2+i\eta}
+2\pi \delta\left(k_0^2-k^2\right)n_\mathrm{B}\left(\vert k_0\vert\right)\right]r^i\,,
\label{transverseleading}
\end{eqnarray}
where $k$ is the modulus of the three-momentum ${\bf k}$. 
Evaluating the integral over the momentum region $k_0,k \sim T \gg (E-h_o^{(0)})$ 
implies that we may expand
\begin{equation}
\label{expan}
\frac{i}{E-h_o^{(0)}-k_0+i\eta}=
\frac{i}{-k_0+i\eta} - i\frac{E-h_o^{(0)}}{(-k_0+i\eta)^2} + i\frac{\left(E-h_o^{(0)}\right)^2}{(-k_0+i\eta)^3}+\ldots\,.
\end{equation}
It is convenient then to regularize the integral in dimensional regularization,
because the non-thermal part, which is scaleless, vanishes, and we are left 
with the thermal part only, which is finite.
The leading contribution comes from the term in \eqref{expan} that is linear in $E-h_o^{(0)}$;  
after performing the integration, it reads
\begin{equation}
\label{linearwithenergy}
\Sigma_s(E)^{\hbox{\tiny Fig.\,\ref{fig:static}}}= 
-\frac{2\pi}{9}\cf\als \left(V_{so}^{(0,1)}(r)\right)^2 T^2\,r^i\left(E-h_o^{(0)}\right)r^i\,.
\end{equation}
A useful identity is 
\begin{equation}
r^i\left(E-h^{(0)}_o\right)r^i  = 
\frac{1}{2}\left[\left[ r^i, E-h^{(0)}_s \right], r^i\right] + \frac{1}{2}\left\{r^2,E-h^{(0)}_s \right\} 
- \left(V_o^{(0)}-V_s^{(0)}\right)r^2\,,
\label{identity}
\end{equation}
which follows from $h^{(0)}_o=h^{(0)}_s+(h^{(0)}_o-h^{(0)}_s)= h^{(0)}_s+(V^{(0)}_o-V^{(0)}_s)$.
The identity is useful, because the first term in the right-hand side is of order $1/m$ and, therefore, 
does not contribute to the static potential, the second term contributes only to the singlet normalization, 
$Z_s$, and the third term to the static potential.
Substituting \eqref{identity} into \eqref{linearwithenergy} and then matching \eqref{hqprops} to \eqref{matchingHTL} 
gives the leading-order thermal corrections to the static potential and the singlet normalization:
\begin{eqnarray}
\delta V_s^{(0)}(r) &=&
\frac{2\pi}{9}\,\cf\als\left(V_{so}^{(0,1)}(r)\right)^2 T^2r^2\left(V^{(0)}_o(r)-V^{(0)}_s(r)\right)\,,
\label{leadingstatic}
\\
\delta Z_s(r) &=& 
-\frac{\pi}{9}\,\cf\als \left(V_{so}^{(0,1)}(r)\right)^2 T^2 r^2 \,,
\label{leadingZs}
\end{eqnarray}
where $V^{(0)}_o(r)-V^{(0)}_s(r) = \nc\als/(2r)$.
The power counting of pNRQCD and Eq. \eqref{hierarchy} give the size of 
$\delta V_s^{(0)}$: $m\als^3 \gg \delta V_s^{(0)} \sim \als^2T^2r\gg m \als^5$.
We recall that at higher orders $\delta V_s^{(0)}$ develops also an imaginary part \cite{Brambilla:2008cx}
(see also \cite{Laine:2006ns}).

\subsection{Singlet spin-orbit potential $\delta V_{LS\,s a}$ in $\mathrm{pNRQCD}_{\rm HTL}$}
In \cite{Brambilla:2010vq}, all contributions to $\delta V_s$, static and non-static, 
that contribute to the spectrum up to order $m\als^5$ were computed. 
Up to that order, no spin-dependent corrections are relevant. 
The aim of this section is to compute the leading spin-orbit terms in $\delta V_s$. 
In particular, we will compute the leading thermal correction, $\delta V_{LS\,s a}$, 
to the centre-of-mass momentum dependent spin-orbit potential $V_{LS\,s a}$, 
defined in Eq. \eqref{defvls}.
The computation follows the same line as the one for $\delta V_s^{(0)}$: 
we calculate thermal spin-dependent corrections to the pNRQCD singlet propagator, 
and match it to the singlet propagator in $\mathrm{pNRQCD}_{\rm HTL}$.

We identify the following set of contributions to $\delta V_{LS\,s a}$:
\begin{equation}
\delta V_{LS\,s a}=\delta V_{LS,a}+\delta V_{LS,b}+\delta V_{LS,c}+\delta V_{LS,d}+\delta V_{LS,e}
\,,
\label{sumP}
\end{equation}
where 
\begin{itemize}
\item[(1)]{$\delta V_{LS,a}$ comes from inserting a spin-orbit potential 
in the singlet or octet propagators of the diagram in Fig.~\ref{fig:static};} 
\item[(2)]{$\delta V_{LS,b}$ comes from replacing one of the two chromoelectric dipole vertices 
in Fig.~\ref{fig:static} with the chromomagnetic vertex proportional 
to $c_FV_{so\,b}^{(1,0)}$ in Eq. \eqref{1/m} and inserting a centre-of-mass kinetic energy 
into the octet propagator;} 
\item[(3)]{$\delta V_{LS,c}$ comes from replacing one of the chromoelectric dipole vertices 
in Fig.~\ref{fig:static} with the vertex proportional to  $c_sV_{so\,a}^{(2,0)}$ in Eq. \eqref{1/m2};}
\item[(4)]{$\delta V_{LS,d}$ comes from replacing one of the chromoelectric dipole vertices 
in  Fig.~\ref{fig:static} with the vertex proportional to  $V_{so\,b''}^{(2,0)}$ in Eq. \eqref{1/m2};} 
\item[(5)]{$\delta V_{LS,e}$ comes from replacing one of the chromoelectric dipole vertices 
in  Fig.~\ref{fig:static} with the vertex proportional to  $c_FV_{so\,b}^{(1,0)}$ in Eq. \eqref{1/m}  
and the other one with the vertex given by $h_{so}^{(1,1)}$ in Eq. \eqref{1/m1m2}.} 
\end{itemize}
By explicit inspection, one sees that diagrams with vertices given by the terms 
proportional to $V^{(1,0)}_{so\,c}$, $V_{so\,b'}^{(2,0)}$ and $V_{so\,b'''}^{(2,0)}(r)$ 
in Eqs. \eqref{1/m} and \eqref{1/m2}, albeit spin-dependent, do not contribute 
to the spin-orbit potential.

\FIGURE[ht]{
\parbox{15cm}{
\centering
\includegraphics[width=15cm]{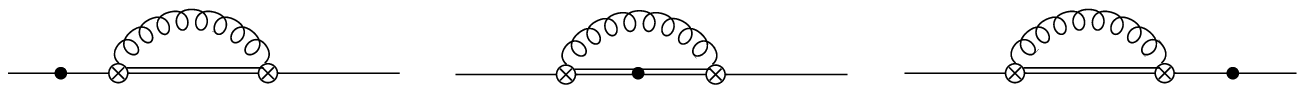}
\caption{Diagrams contributing to $\delta V_{LS,a}$. 
The dot stands for an insertion of the spin-orbit potential proportional to $V_{LS\,sa}$
(left and right diagram) or to $V_{LS\,oa}$ (middle diagram), 
all other symbols are as in Fig.~\ref{fig:static}.}
\label{diagram1}}
}

\subsubsection{Evaluation of $\delta V_{LS,a}$}
We evaluate the diagrams in Fig.~\ref{diagram1}.
As in the previous calculation of the thermal correction to the static propagator, 
we expand the octet propagators for $k_0\gg E-h_o^{(0)}$ (see Eq. \eqref{expan}). 
The leading contribution comes again from the linear term that we treat by means 
of the identity \eqref{identity}. The left diagram of Fig.~\ref{diagram1} gives
\begin{eqnarray}
&& \hspace{-9mm}
-\frac{2\pi}{9}C_F\als T^2
\frac{i}{E-h_s^{(0)}} 
\left(V_{so}^{(0,1)}(r)\right)^2
\left[ \frac{1}{2}\left\{r^2,E-h^{(0)}_s \right\} - \left(V_o^{(0)}(r)-V_s^{(0)}(r)\right)r^2 \right]
\nonumber\\
&& \times
\frac{1}{E-h_s^{(0)}} \frac{({\bf r}\times {\bf P}) \cdot (\bfsigma^{(1)} - \bfsigma^{(2)})}{4 m^2} V_{LS\,s a}(r)
\frac{1}{E-h_s^{(0)}}\,,
\label{left}
\end{eqnarray}
the right one gives
\begin{eqnarray}
&& \hspace{-9mm}
-\frac{2\pi}{9}C_F\als T^2
\frac{i}{E-h_s^{(0)}} \frac{({\bf r}\times {\bf P}) \cdot (\bfsigma^{(1)} - \bfsigma^{(2)})}{4 m^2} V_{LS\,s a}(r)
\frac{1}{E-h_s^{(0)}} 
\nonumber\\
&& \times
 \left(V_{so}^{(0,1)}(r)\right)^2
\left[ \frac{1}{2}\left\{r^2,E-h^{(0)}_s \right\} - \left(V_o^{(0)}(r)-V_s^{(0)}(r)\right)r^2 \right] 
\frac{1}{E-h_s^{(0)}}\,,
\label{right}
\end{eqnarray}
and the middle one gives
\begin{eqnarray}
&& \hspace{-9mm}
\frac{2\pi}{9}C_F\als T^2
\frac{i}{E-h_s^{(0)}} 
 \left(V_{so}^{(0,1)}(r)\right)^2 r^2
\frac{({\bf r}\times {\bf P}) \cdot (\bfsigma^{(1)} - \bfsigma^{(2)})}{4 m^2} V_{LS\,o a}(r)
\frac{1}{E-h_s^{(0)}}\,,
\nonumber\\
\label{middle}
\end{eqnarray}
where we have kept only terms relevant at order $1/m^2$.
Matching to the $\textrm{pNRQCD}_\mathrm{HTL}$ propagator \eqref{matchingHTL}, 
expanded according to \eqref{hqpropsozero}, we observe that 
the terms proportional to $(V_o^{(0)}-V_s^{(0)})r^2$ in \eqref{left} and \eqref{right} 
cancel against one insertion of $\delta V_s^{(0)}(r)$ and one of the spin-orbit potential in 
the  $\textrm{pNRQCD}_\mathrm{HTL}$ propagator, while the term proportional to 
$(E-h^{(0)}_s) r^2/2$ in \eqref{left} and the one proportional to 
$r^2 (E-h^{(0)}_s)/2$ in \eqref{right} cancel against the term 
$\left\{ \delta Z_s, i/(E-h^{(0)}_s)\right.$ 
$\times [\hbox{spin-orbit potential}]$ $ \times$ $\left.1/(E-h^{(0)}_s) \right\}$
in  \eqref{matchingHTL}. The expression of $\delta Z_s$ can be read from Eq. \eqref{leadingZs}.
What is left gives the leading-order thermal correction, coming from the diagrams in Fig.~\ref{diagram1}, 
to the centre-of-mass  momentum dependent spin-orbit potential:
\begin{equation}
\delta V_{LS,a}(r)=\frac{2\pi}{9}C_F\als \left(V_{so}^{(0,1)}(r)\right)^2 T^2r^2\left(V_{LS\,oa}(r)-V_{LS\,sa}(r)\right)\,.
\label{aspinorbit}
\end{equation}
According to the power counting of pNRQCD and Eq. \eqref{hierarchy}, we have that 
$m\als^5\gg \delta V_{LS,a}\,({\bf r}\times {\bf P})\cdot
(\bfsigma^{(1)} - \bfsigma^{(2)})/m^2 \gg m \als^8$.

\FIGURE[ht]{
\parbox{15cm}{
\centering
\includegraphics[width=12cm]{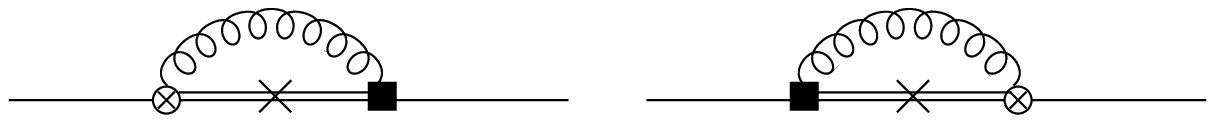} 
\caption{Diagrams contributing to $\delta V_{LS,b}$. 
The square stands for the chromomagnetic vertex proportional to $c_FV_{so\,b}^{(1,0)}(r)$, 
the cross for a centre-of-mass kinetic energy insertion, and 
all other symbols are as in Fig.~\ref{fig:static}.}
\label{diagram3}}
}

\subsubsection{Evaluation of $\delta V_{LS,b}$}
We evaluate the diagrams in Fig.~\ref{diagram3}. Their thermal contribution to $\Sigma_s(E)$ reads 
\begin{eqnarray}
\Sigma_s(E)^{\hbox{\tiny Fig.\,\ref{diagram3}}}&=&
- 2 ig^2\cf V_{so}^{(0,1)}(r)V_{so\,b}^{(1,0)}(r)\frac{c_F}{2m}
\nonumber\\
&& \hspace{-1cm}\times
r^i \int\frac{\,d^4k}{(2\pi)^4}
\frac{i}{E-h_o^{(0)}-k_0+i\eta}
\frac{({\bf P}-{\bf k})^2}{4m}
\frac{1}{E-h_o^{(0)}-k_0+i\eta}
\nonumber\\
&&\hspace{-1cm}\times  
k_0\,\epsilon^{jln} k^l 
\left(\delta^{ni}-\frac{k^nk^i}{k^2}\right)\,2\pi\delta(k_0^2-k^2)
n_\mathrm{B}(\vert k_0\vert)\,
(\sigma^{(1)\,j}-\sigma^{(2)\,j})
\,,
\label{sigmab}
\end{eqnarray}
while the non-thermal part vanishes if regularized in dimensional regularization.
The factor $2$ follows from the fact that the two diagrams give the same contribution 
at order $1/m^2$. The octet propagators may be expanded according to Eq. \eqref{expan}; 
considering that, besides the two octet propagators, the integral in \eqref{sigmab} is 
odd in $k_0$, the leading non-vanishing term coming from their expansion is $-2(E-h_o^{(0)})/(-k_0+i\eta)^3$.
The factor $E-h_o^{(0)}$ contains a part, $E-h_s^{(0)}$, that contributes to the singlet normalization, 
and a part, $V_s^{(0)}-V_o^{(0)}$, that contributes to the spin-orbit potential.
The octet centre-of-mass kinetic energy, ${({\bf P}-{\bf k})^2}/(4m)$, contributes to the 
spin-orbit potential only through the term  $- {\bf P}\cdot{\bf k}/(2m)$.
With this in mind, we match \eqref{hqprops} to \eqref{matchingHTL} and obtain
the leading-order thermal correction, coming from the diagrams in Fig.~\ref{diagram3}, 
to the centre-of-mass  momentum dependent spin-orbit potential:
\begin{equation}
\delta V_{LS,b}(r) = 
-\frac{4\pi}{9}C_F \als  c_F V_{so}^{(0,1)}(r) V_{so\,b}^{(1,0)}(r) T^2 \left(V^{(0)}_o(r)-V^{(0)}_s(r)\right)\,.
\label{bspinorbit}
\end{equation}
Considering that the matching coefficients $c_F$, $V_{so}^{(0,1)}$ and  $V_{so\,b}^{(1,0)}$ are one 
at leading order, the size of the correction is 
$m\als^5\gg \delta V_{LS,b}\,({\bf r}\times {\bf P})\cdot (\bfsigma^{(1)} - \bfsigma^{(2)})/m^2 \gg m\als^8$.

\subsubsection{Evaluation of $\delta V_{LS,c}$}
\label{subsecc}
The calculation of $\delta V_{LS,c}$ is at this point simple: 
there are two contributing diagrams, which may be constructed 
by replacing one of the chromoelectric dipole vertices 
in Fig.~\ref{fig:static} with the vertex proportional 
to  $c_sV_{so\,a}^{(2,0)}$ in Eq. \eqref{1/m2}.
Since this vertex contains a chromoelectric field as well, the integration is exactly the
same as the one performed in Eq. \eqref{transverseleading}. The only
change is in the prefactor of the integral. Matching to the 
$\textrm{pNRQCD}_\mathrm{HTL}$ propagator, we obtain at leading order 
\begin{equation}
\delta V_{LS,c}(r) = \frac{2\pi}{9}C_F\als c_sV_{so}^{(0,1)}(r) V_{so\,a}^{(2,0)}(r) T^2\left(V^{(0)}_o(r)-V^{(0)}_s(r)\right)\,,
\label{cspinorbit}
\end{equation}
which, considering that  $c_s$, $V_{so}^{(0,1)}$ and $V_{so\,a}^{(2,0)}$ are one at leading order, 
has the same size as $\delta V_{LS,b}$.

\subsubsection{Evaluation of $\delta V_{LS,d}$}
The calculation of $\delta V_{LS,d}$ is similar to this last one, but with the vertices 
proportional to  $c_sV_{so\,a}^{(2,0)}$ replaced by the ones proportional to $V_{so\,b''}^{(2,0)}$ in Eq. \eqref{1/m2}.
The leading-order result reads
\begin{equation}
\delta V_{LS,d}(r) = \frac{2\pi}{9}C_F\als V_{so}^{(0,1)}(r) V_{so\,b''}^{(2,0)}(r) T^2 \left(V^{(0)}_o(r)-V^{(0)}_s(r)\right)\,.
\label{dspinorbit}
\end{equation}
Considering that $V_{so}^{(0,1)}$ is one at leading order, but that $V_{so\,b''}^{(2,0)}$ is 
at least of order $\als^3$, $\delta V_{LS,d}(r)$ is suppressed with respect to  
$\delta V_{LS,a}$, $\delta V_{LS,b}$ and $\delta V_{LS,c}$ by, at least, a factor $\als^3$.

\FIGURE[ht]{
\parbox{15cm}{
\centering
\includegraphics[width=12cm]{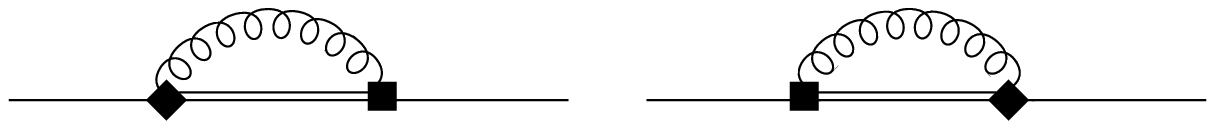} 
\caption{Diagrams contributing to $\delta V_{LS,e}$. 
The diamond stands for the vertex $h_{so}^{(1,1)}$, given in Eq. \eqref{1/m1m2}, 
and all other symbols are as in Fig.~\ref{diagram3}.}
\label{diagram4}}
}

\subsubsection{Evaluation of $\delta V_{LS,e}$}
We evaluate the diagrams in Fig.~\ref{diagram4}. Their thermal contribution to $\Sigma_s(E)$ reads 
\begin{eqnarray}
\Sigma_s(E)^{\hbox{\tiny Fig.\,\ref{diagram4}}}&=&
- 2ig^2\cf V_{so}^{(1,1)}(r)V_{so\,b}^{(1,0)}(r)\frac{c_F}{2m}
\left(- \frac{({\bf P}\times{\bf r})^i}{2m} \right)
\int\frac{\,d^4k}{(2\pi)^4}
\frac{i}{E-h_o^{(0)}-k_0+i\eta}
\nonumber\\
&& \hspace{-1cm}\times
(ik^l)\epsilon^{jlr} \, \epsilon^{ins} (-ik^n) 
\left(\delta^{rs}-\frac{k^rk^s}{k^2}\right)\,2\pi\delta(k_0^2-k^2)
n_\mathrm{B}(\vert k_0\vert)\,
(\sigma^{(1)\,j}-\sigma^{(2)\,j})
\,,
\label{sigmae}
\end{eqnarray}
while the non-thermal part vanishes if regularized in dimensional regularization.
The factor $2$ follows from the fact that the two diagrams give the same contribution 
at order $1/m^2$. The octet propagators may be expanded according to Eq. \eqref{expan}: 
the linear term in $E-h_o^{(0)}$ contains a part, $E-h_s^{(0)}$, that contributes to the singlet normalization, 
and a part, $V_s^{(0)}-V_o^{(0)}$, that contributes to the spin-orbit potential.
This last contribution reads
\begin{equation}
\delta V_{LS,e}(r) = 
\frac{4\pi}{9}C_F \als  c_F V_{so}^{(1,1)}(r) V_{so\,b}^{(1,0)}(r) T^2 \left(V^{(0)}_o(r)-V^{(0)}_s(r)\right)\,.
\label{espinorbit}
\end{equation}
Considering that, according to \eqref{poincare11}, 
the matching coefficient $V_{so}^{(1,1)}$  is equal to $V_{so}^{(0,1)}$, 
$\delta V_{LS,e}$ exactly cancels with $\delta V_{LS,b}$ in the sum \eqref{sumP}.

\subsubsection{Summary}
In summary, the leading thermal correction to the centre-of-mass momentum-depen\-dent 
spin-orbit potential,  
\begin{equation}
\delta V_{LS\,s}|_{{\bf P}\hbox{\tiny -dependent}} = 
\frac{({\bf r}\times {\bf P}) \cdot (\bfsigma^{(1)} - \bfsigma^{(2)})}{4 m^2} \delta V_{LS\,s a}(r)\,,
\end{equation}
is the sum of Eqs. \eqref{aspinorbit}, \eqref{bspinorbit}, \eqref{cspinorbit}, \eqref{dspinorbit}
and  \eqref{espinorbit}; it reads:
\begin{eqnarray}
\hspace{-7mm}
\delta V_{LS\,s a}(r) &=& \frac{2\pi}{9}C_F\als V_{so}^{(0,1)}(r) T^2
\left\{
V_{so}^{(0,1)}(r) r^2\left(V_{LS\,oa}(r)-V_{LS\,sa}(r)\right) \right.
\nonumber\\
&& 
\left. +
\left[ c_sV_{so\,a}^{(2,0)}(r) + V_{so\,b''}^{(2,0)}(r)       \right] 
\left(V^{(0)}_o(r)-V^{(0)}_s(r)\right) 
\right\}
\nonumber\\
&=& \frac{\pi}{6}\cf N_c \frac{\als^2}{r}T^2 + \hbox{higher orders}\,.
\label{summaryLS}
\end{eqnarray}
In the first equality of \eqref{summaryLS}, the matching coefficients of NRQCD and pNRQCD have been kept unexpanded; 
this amounts at having provided an expression for the spin-orbit potential that resums 
contributions from the scales $m$ and  $m\als$, while it is of leading order in the temperature.
In the second equality, we have kept only the leading terms in the NRQCD and pNRQCD matching coefficients.
We note that the contribution coming from the term proportional to $V_{so\,b''}^{(2,0)}$ is negligible, 
of the same size or smaller than subleading thermal corrections that we have neglected throughout the paper.

\section{Gromes relation at finite temperature}
\label{secpoi}
After having computed the leading contributions to $\delta V_{LS\,sa}$, 
we can now check whether these new terms fulfill the Gromes relation \eqref{gromes} or not. 
This corresponds to verifying the equality
\begin{equation}
\label{testgromes}
\delta V_{LS\,sa}(r) \stackrel{?}{=} - \frac{\delta V^{(0)}_{s}(r)^\prime}{2r} \,.
\end{equation}
We use the expression of $\delta V_{LS\,sa}$ provided by the first equality in Eq. \eqref{summaryLS}
that keeps unexpanded  the matching coefficients of NRQCD and pNRQCD. 
If we make use of the relations  \eqref{gromes} and \eqref{gromesmatching1}, which are exact, 
then $\delta V_{LS\,sa}$ may be rewritten as 
\begin{eqnarray}
\delta V_{LS\,s a}(r) &=&
- \frac{\delta V^{(0)}_{s}(r)^\prime}{2r} 
\nonumber\\
&& 
+ \frac{2\pi}{9}C_F\als V_{so}^{(0,1)}(r) T^2
\left( c_sV_{so\,a}^{(2,0)}(r) + V_{so}^{(0,1)}(r)\right)
\left(V^{(0)}_o(r) - V^{(0)}_s(r)\right),
\label{nocancel}
\end{eqnarray}
which shows that the Gromes relation is violated by an amount, which at leading order is  
$\displaystyle \frac{2\pi}{9}\cf N_c \frac{\als^2}{r}T^2$.

\subsection{The spin-orbit potential in a covariant model}
In order to understand the origin of the observed violation of Poincar\'e invariance, it 
is useful to consider at zero temperature the case of a massive gluon, whose mass, $m_g$, is such that 
$m\als \gg m_g \gg m\als^2$. The massive gluon contributes to the potential, but 
clearly it does not break  Poincar\'e invariance. To see this let us evaluate the 
corrections to the spin-orbit potential. 
The diagrams contributing to the spin-orbit potential are the same of those 
considered in the previous section, only now the gluon propagator reads (in the unitary gauge)
\begin{equation}
-\frac{i}{k_0^2-k^2-m_g^2+i\eta}\left(g_{\mu\nu} - \frac{k_\mu k_\nu}{m_g^2}\right).
\end{equation}

The contributions to the static potential, $\delta V_s^{(0)}$, and 
 $\delta V_{LS,a}$, $\delta V_{LS,c}$ and $\delta V_{LS,d}$  depend on the 
correlator of two chromoelectric fields. They are proportional to 
\begin{equation}
\int\frac{\,d^dk}{(2\pi)^d}\frac{i}{E-h_o^{(0)}-k_0+i\eta} 
\frac{1}{k_0^2-k^2-m_g^2+i\eta} \left[(d-1)\,k_0^2-k^2\right],
\end{equation} 
where we have regularized the integral in dimensional regularization.
Integrating over the momentum region $k_0, k \sim m_g$ means that we are expanding the octet 
propagator according to Eq. \eqref{expan}. We will focus here only on the term linear in 
$E-h_o^{(0)}$, since this is the relevant term in the finite temperature case analyzed  
in the rest of the paper.\footnote{
We have explicitly checked that also the leading contribution to the spin-orbit potential, proportional to 
$m_g^3/m^2$, satisfies the Gromes relation \cite{jacopo}.
}
It turns out that the linear term vanishes in dimensional regularization (see \cite{Brambilla:1999xf}). 
The reason is that the contribution coming from the spatial components 
of the gluon propagator (proportional to $(d-1)\,k_0^2$ in the equation)
cancels against the contribution coming from the temporal component (proportional to $k^2$ in the equation). 
This is in sharp contrast with the finite temperature case, where the term linear in $E-h_o^{(0)}$ does not vanish
(in Coulomb gauge, this is due to the fact that only the spatial components of the gluon propagator get thermal 
contributions) and eventually generate a finite thermal contribution 
to $\delta V_s^{(0)}$, $\delta V_{LS,a}$, $\delta V_{LS,c}$ and $\delta V_{LS,d}$ 
(see Eqs. \eqref{leadingstatic}, \eqref{aspinorbit}, \eqref{cspinorbit} and \eqref{dspinorbit}
respectively).

The contributions to  $\delta V_{LS,b}$ and $\delta V_{LS,e}$ depend on the spatial components 
of the gluon propagator only. Both $\delta V_{LS,b}$ and $\delta V_{LS,e}$ get finite 
contributions from the massive gluon but the sum of the two terms linear in $E-h_o^{(0)}$
vanishes: the same happens in the finite temperature case discussed in the previous section.

The massive gluon example provides a simple case where Poincar\'e invariance is not broken.
The Gromes relation is trivially realized for terms that are linear in the energy: 
such terms vanish for both the static and the spin-orbit potentials.  
In the finite temperature case, diagrams that depend on the 
correlator of two chromoelectric fields,  like the one shown in Fig.~\ref{fig:static}, 
do not vanish. This is a direct consequence of the fact that the thermal bath 
affects in a non-covariant way the gluon propagator.

\section{Singlet spin-orbit potential $\delta V_{LS\,s b}$ in $\mathrm{pNRQCD}_{\rm HTL}$}
\label{secspinorb}
In this section, we calculate the leading thermal corrections to the spin-orbit potential 
$\delta V_{LS\,sb}$, which is the spin-orbit potential experienced by the quarkonium when at rest with 
respect to the laboratory reference frame (we recall that, in our setup, this is also the reference frame of the 
thermal bath). This potential, even at zero temperature, is not constrained by Poincar\'e invariance.

In order to calculate $\delta V_{LS\,sb}$, we need to consider two new terms contributing to $h_{so}$ 
in the pNRQCD Lagrangian: the term 
\begin{equation}
- \frac{c_F}{4m} \, \bfsigma^{(1)}\cdot r^i (\partial_i\,g{\bf B})\,,
\label{so2}
\end{equation}
and the term
\begin{equation}
\frac{c_s}{8m^2} \, \bfsigma^{(1)}\cdot [{\bf p} \times, g {\bf E}]\,,
\label{so1}
\end{equation}
where, for simplicity, we have put to their tree-level values the pNRQCD matching coefficients.

There are three classes of diagrams that contribute: 
\begin{equation}
\delta V_{LS\,sb} = \delta V_{LS\,(i)} + \delta V_{LS\,(ii)} + \delta V_{LS\,(iii)}.
\end{equation}
\begin{itemize}
\item[(1)]{The first class is similar to the one shown in Fig.~\ref{diagram1}, but now the 
dots stand for insertions of the spin-orbit potential proportional to $V_{LS\,sb}$
(left and right diagram) or to $V_{LS\,ob}$ (middle diagram). $V_{LS\,sb}$ and 
$V_{LS\,ob}$ have been defined in Eqs. \eqref{defvls} and \eqref{defvlso}, 
and given at leading order in Eq. \eqref{spinorbitlo}. The result reads
\begin{equation}
\delta V_{LS\,(i)}(r)=\frac{2\pi}{9}C_F\als T^2r^2\left(V_{LS\,ob}(r)-V_{LS\,sb}(r)\right)\,.
\label{aprimespinorbit}
\end{equation}
}
\item[(2)]{The second class is similar to the one shown in Fig.~\ref{diagram3}, 
but now the squares stand for the vertex induced by \eqref{so2} and the cross for a 
kinetic energy insertion, ${\bf p}^2/m$. The result reads
\begin{equation}
\delta V_{LS\,(ii)}(r) = 
-\frac{4\pi}{9}C_F \als  c_F T^2 \left(V^{(0)}_o(r)-V^{(0)}_s(r)\right)\,.
\label{bprimespinorbit}
\end{equation}
}
\item[(3)]{Finally, the third class of diagrams is similar to the ones evaluated in 
Sec.~\ref{subsecc}, but with the vertex proportional to $c_sV_{so\,a}^{(2,0)}$ in Eq. \eqref{1/m2} 
replaced by the vertex induced by \eqref{so1}. The result reads 
\begin{equation}
\delta V_{LS\,(iii)}(r) = 
\frac{2\pi}{9}C_F \als  c_s T^2 \left(V^{(0)}_o(r)-V^{(0)}_s(r)\right)\,.
\label{cprimespinorbit}
\end{equation}
}
\end{itemize}

Summing up all three contributions we obtain
\begin{eqnarray}
\hspace{-7mm}
\delta V_{LS\,s b}(r) &=& \frac{2\pi}{9}C_F\als T^2
\left[r^2\left(V_{LS\,ob}(r)-V_{LS\,sb}(r)\right) 
- \left(V^{(0)}_o(r)-V^{(0)}_s(r)\right) 
\right]
\nonumber\\
&=& - \frac{5\pi}{18}\cf N_c \frac{\als^2}{r}T^2 + \hbox{higher orders}\,,
\label{summaryLScm}
\end{eqnarray}
where, in the first equality, we have used that $2c_F-c_s-1=0$.

\section{Conclusions}
\label{secconclusions}
We have calculated the leading-order thermal corrections to the quarkonium spin-orbit 
potentials. These corrections go quadratically with the temperature and are proportional to 
$\als^2T^2/r$.

At zero temperature, the spin-orbit potential that depends on the centre-of-mass momentum is 
protected by Poincar\'e invariance. We have computed its leading thermal correction 
in Eq. \eqref{summaryLS}. In Eq. \eqref{nocancel}, this correction has been shown 
to violate Poincar\'e invariance. This implies that order $\als^2T^2/r$ corrections to the
quarkonium potential will be experienced by the system differently in different 
reference frames, and, in particular, in a frame where the thermal bath is not at rest.

We have also computed the leading thermal correction to the spin-orbit potential 
of a quarkonium at rest with respect to the laboratory reference frame.
Its expression is in Eq. \eqref{summaryLScm}.
The potential contributes to the spin-orbit splittings of the quarkonium levels. 
The thermal correction being negative implies a weakening of the spin-orbit interaction in the medium. 

In the special case of a bound state made of two heavy quarks with different masses, 
$m_1 \gg m_2$ (like the $B_c$ system), the complete spin-dependent potential, 
up to corrections of relative order $m_2/m_1$, is given by 
$(V_{LS\,s a}+V_{LS\,s b})({\bf r}\times{\bf p}_2)\cdot \bfsigma_2/(4m_2^2)$.
Therefore, the leading contribution to the hyperfine splittings reads
\begin{equation}
\delta E_{njl} = 
- \frac{\pi}{36}\cf^2 N_c \frac{\als^3}{m_2}T^2 \frac{j(j+1)-l(l+1)-3/4}{n^2}\,,
\end{equation}
where $n$, $j$ and $l$ are the principal, orbital and total angular momentum quantum numbers.
The thermal correction has opposite sign with respect to the splitting at zero temperature, 
which implies smaller hyperfine splittings in the medium.

\acknowledgments
M.A.E. acknowledges useful discussions with Joan Soto in the early stages of this work.
We acknowledge financial support from the DFG project BR4058/1-1 
``Effective field theories for strong interactions with heavy quarks''.
N.B., J.G. and A.V.  acknowledge financial support 
from the DFG cluster of excellence ``Origin and structure of the universe'' 
(\href{http://www.universe-cluster.de}{www.universe-cluster.de}). 
M.A.E. acknowledges financial support from the RTN Flavianet MRTN-CT-2006-035482 (EU)
and the University of Milano for hospitality during the early stages of this work.

\end{document}